\documentclass[12pt]{article}
\usepackage{amssymb,amsfonts}
\usepackage{epsf,epsfig}
\def \di{\displaystyle}

\begin{document}
\baselineskip 18pt

\title{Magnon mode truncation in a rung-dimerized asymmetric spin ladder}
\author{P.~N.~Bibikov\\ \it V.~A.~Fock Institute of Physics
Sankt-Petersburg State University}

\maketitle

\vskip5mm

\begin{abstract}
An effective model is suggested for an asymmetric spin ladder with
dimerized rungs. Magnon mode truncation originated from magnon
decay (recently observed in the 1D compound ${\rm IPA-CuCl}_3$) is
naturally described within this model. Using Bethe Ansatze we
studied a one-magnon sector and obtained relations between
interaction constants of the model and experimentally observable
quantities such as the gap and truncation energies, spin velocity
and truncation wave vector. It is also shown that the structure
factor turns to zero at the truncation point.
\end{abstract}

\section{Introduction}

A spin ladder with strong antiferromagnetic rung coupling gives an
ideal example of a gapped spin-dimerized system \cite{1}. Really
the majority of spins in its ground state are coupled in
rung-singlets (rung-dimers) so that the relative coupling energy
estimates a value of the gap. By this reason all low-temperature
effects depend on dynamical properties of states with a few number
of excited rungs. Theoretical study of excitations in spin ladders
with strong antiferromagnetic rung coupling was developed in a
number of papers \cite{2}-\cite{5}. It was pointed out that the
lowest excitations form a coherent magnon branch. When the gap
energy is smaller than the energy width of the magnon zone the
latter may intersect the two-magnon scattering continuum. For a
{\it symmetric} spin ladder (with equal couplings along both legs
as well as along both diagonals) these two sectors do not
hybridize so a one-magnon state is always stable. The situation is
quite different for an {\it asymmetric} spin ladder with non equal
couplings along legs or along diagonals. As it was pointed in
\cite{6}-\cite{9} the coupling asymmetry entails hybridization
between the "bare" (related to a symmetric case) one- and
two-magnon sectors. If the system has a wide magnon band
intersecting with the two-magnon scattering continuum this
hybridization results to magnon instability and truncation of the
magnon mode at some value $k_{trunc}$ of wave vector.
Experimentally a magnon mode truncation was observed in neuron
scattering from 1D compound ${\rm IPA-CuCl}_3$ (${\rm(CH}_3)_2{\rm
CHNH}_3{\rm CuCl}_3$) \cite{6}. The latter is considered as an
asymmetric spin ladder with strong ferromagnetic rungs and is
effectively equivalent to a 1D Haldane antiferromagnet.

In this paper starting from an asymmetric rung-dimerized spin
ladder we present an effective model which produce an explicit
realization of magnon mode truncation related to magnon decay.
Within our model we study the one-magnon excitations and obtain
explicit relations between the coupling constants and
experimentally observable quantities.

\section{Hamiltonian for an asymmetric spin ladder}
In the present paper we shall study an asymmetric spin ladder with
the following Hamiltonian \cite{4},\cite{9},
\begin{equation}
{\hat H}=\sum_nH_{n,n+1},
\end{equation}
where
$H_{n,n+1}=H^{rung}_{n,n+1}+H^{leg}_{n,n+1}+H^{diag}_{n,n+1}+H^{cyc}_{n,n+1}+H^{norm}_{n,n+1}$
and
\begin{eqnarray}
H^{rung}_{n,n+1}&=&\frac{J_{\bot}}{2}({\bf S}_{1,n}\cdot{\bf
S}_{2,n}+{\bf S}_{1,n+1}\cdot{\bf
S}_{2,n+1}),\nonumber\\
H^{leg}_{n,n+1}&=&J_{\|}({\bf S}_{1,n}\cdot{\bf S}_{1,n+1}+{\bf
S}_{2,n}\cdot{\bf
S}_{2,n+1}),\nonumber\\
H^{diag}_{n,n+1}&=&J_d{\bf S}_{1,n}\cdot{\bf
S}_{2,n+1},\nonumber\\
H^{cyc}_{n,n+1}&=&J_{c}(({\bf S}_{1,n}\cdot{\bf S}_{1,n+1})({\bf
S}_{2,n}\cdot{\bf S}_{2,n+1})+({\bf S}_{1,n}\cdot{\bf S
}_{2,n})\nonumber\\
&\times&({\bf S}_{1,n+1}\cdot{\bf S}_{2,n+1}) -({\bf
S}_{1,n}\cdot{\bf S}_{2,n+1})({\bf S}_{2,n}\cdot{\bf S}_{1,n+1})).
\end{eqnarray}
Here ${\bf S}_{j,n}$ ($j=1,2$) are the $S=1/2$ spin operators
related to $n$-th rung. The auxiliary term $H^{norm}=J_{norm}I$
($I$ is an identity matrix) is added for the zero normalization of
the ground state energy.

The following condition,
\begin{equation}
J_d+J_c=2J_{\|},
\end{equation}
suggested in \cite{4}, guarantees that the vector,
$|0\rangle_n\otimes|0\rangle_{n+1}$ (where $|0\rangle_n$ is the
$n$-th rung-singlet or equivalently rung-dimer) is an eigenstate
for $H_{n,n+1}$, so the vector
\begin{equation}
|0\rangle=\prod_n|0\rangle_n,
\end{equation}
is an eigenstate for $\hat H$. An additional system of
inequalities,
\begin{eqnarray}
J_{\bot}&>&2J_{\|},\quad J_{\bot}>\frac{5}{2}J_c,\quad J_{\bot}+J_{||}>\frac{3}{4}J_c,\nonumber\\
3J_{\bot}-2J_{\|}-J_c&>&\sqrt{J_{\bot}^2-4J_{\bot}J_{\|}+20J_{\|}^2-16J_{\|}J_c+4J_c^2},
\end{eqnarray}
together with a condition $J_{norm}=3/4J_{\bot}-9/16J_c$,
guarantee that the vector (4) is the (zero energy) ground state
for $\hat H$. The full system of the "ground state tuning"
conditions (3),(5) belongs to the mathematical basis of our model.

For $J_d=0$ the Hamiltonian $\hat H$ commutes with the operator,
$\hat Q=\frac{1}{2}\sum_n({\bf S}_{1,n}+{\bf S}_{2,n})^2$,
considered as a number of "bare" magnons \cite{5},\cite{9}.
Therefore the Hilbert space splits on an infinite sum: ${\cal
H}=\sum_{m=0}^{\infty}{\cal H}^m$, where $\hat Q|_{{\cal H}^m}=m$.
The subspace ${\cal H}^0$ is one-dimensional and generated by the
ground state (4).

\section{Spectral problem for the reduced Hamiltonian related to the effective model}

Despite the ground state (4) for the Hamiltonian (1)-(3), (5) is
known it is not clear how to obtain its excitations. In the
symmetric case \cite{4},\cite{5} the one-magnon state corresponds
to ${\cal H}^1$ but even for a small asymmetry it already lies in
$\sum_{n=0}^{\infty}{\cal H}^{2n+1}$ \cite{9}. By this reason the
related spectral problem seems to be unsolvable. However for a
strong rung coupling the states with rather big number of "bare"
magnons have a large energy and therefore may be effectively
reduced. In the first order with respect to the dimerization
energy the reduced Hilbert space ${\cal H}^{red}={\cal
H}^0\oplus{\cal H}^1\oplus{\cal H}^2$ contains additionally to the
ground state (4), only the "bare" one- and two-magnon sectors. The
corresponding effective Hamiltonian $\hat H^{eff}$ is defined as
the restriction of $\hat H$ on ${\cal H}^{red}$ or,
\begin{equation}
{\hat H}^{eff}=P^{(0,1,2)}{\hat H}P^{(0,1,2)},
\end{equation}
where $P^{(0,1,2)}$ is the projector on ${\cal H}^{red}$.

A general $S=1$ excited state for $\hat H^{eff}$ related to a wave
vector $k$ and the energy $E(k)$ is superposition of "bare" one-
and two-magnon components,
\begin{equation}
|k\rangle^{\alpha}=\frac{1}{Z(k)\sqrt{N}}\sum_m[a(k){\rm e}^{ikm}...|1\rangle^{\alpha}_m...+
\varepsilon_{\alpha\beta\gamma}\sum_{n>m}{\rm e}^{ik(m+n)/2}b(k,n-m)
...|1\rangle^{\beta}_m...|1\rangle^{\gamma}_n...],
\end{equation}
where $|1\rangle_n^{\alpha}=({\bf S}^{\alpha}_{1,n}-{\bf
S}^{\alpha}_{2,n})|0\rangle_n$ and "..." means an infinite product
of dimers related to the remaining rungs. The normalization factor
$Z(k)$ is defined as,
\begin{equation}
Z^2(k)=|a(k)|^2+2\sum_{n=1}^{\infty}|b(k,n)|^2.
\end{equation}

The system of ${\rm Shr\ddot odinger}$ equations on the amplitudes $a(k)$ and $b(k,n)$ directly follows from the local
action of the operator $H_{n,n+1}$,
\begin{eqnarray}
H_{n,n+1}|0\rangle_n|1\rangle_{n+1}^{\alpha}&=&(\frac{1}{2}J_{\bot}-\frac{3}{4}J_c)|0\rangle_n|1\rangle_{n+1}^{\alpha}+
\frac{J_c}{2}|1\rangle_n^{\alpha}|0\rangle_{n+1}-\frac{iJ_d}{2}
\varepsilon_{\alpha\beta\gamma}|1\rangle_n^{\beta}|1\rangle_{n+1}^{\gamma},\nonumber\\
H_{n,n+1}|1\rangle_n^{\alpha}|0\rangle_{n+1}&=&(\frac{1}{2}J_{\bot}-\frac{3}{4}J_c)|1\rangle_n^{\alpha}|0\rangle_{n+1}+
\frac{J_c}{2}|0\rangle_n|1\rangle_{n+1}^{\alpha},\nonumber\\
H_{n,n+1}\varepsilon_{\alpha\beta\gamma}|1\rangle_n^{\beta}|1\rangle_{n+1}^{\gamma}&=&
(J_{\bot}-J_{\|}-J_c/4)\varepsilon_{\alpha\beta\gamma}
|1\rangle_n^{\beta}|1\rangle_{n+1}^{\gamma}+iJ_d|0\rangle_n|1\rangle_{n+1}^{\alpha}.
\end{eqnarray}
From (7) and (9) one can obtain an infinite set of recurrent
equations,
\begin{equation}
(2J_{\bot}-3J_c)b(k,n)+J_c\cos\frac{k}{2}[b(k,n-1)+b(k,n+1)]=E(k)b(k,n),\quad n>1,
\end{equation}
related to non- neighbor excited rungs and two additional
equations related to neighbor rungs,
\begin{eqnarray}
(J_{\bot}-\frac{3}{2}J_c+J_c\cos k)a(k)+iJ_d\cos\frac{k}{2}b(k,1)&=&E(k)a(k),\nonumber\\
(2J_{\bot}-\frac{9}{4}J_c-\frac{J_d}{2})b(k,1)+J_c\cos\frac{k}{2}b(k,2)-\frac{iJ_d}{2}\cos\frac{k}{2}a(k)&=&
E(k)b(k,1).
\end{eqnarray}

For a coherent excitation originated from the hybridization of the
one-magnon and bound two-magnon states there must be,
\begin{equation}
\lim_{n\rightarrow\infty}b(k,n)=0.
\end{equation}
With regard to this condition the Eq. (10) has the following
general solution,
\begin{equation}
b(k,n)=B(k)z^n(k),
\end{equation}
where,
\begin{equation}
|z(k)|<1.
\end{equation}
and
\begin{equation}
E(k)=2J_{\bot}-3J_c+J_c\Big(z(k)+\frac{1}{z(k)}\Big)\cos\frac{k}{2}.
\end{equation}
From (14) and (15) follows that,
\begin{equation}
{\rm Im}\,z(k)=0.
\end{equation}

Substituting (13) and (15) into (11) we obtain a pair of equations
on $a(k)$ and $B(k)$ represented in the following matrix form,
\begin{equation}
M(k)\left(\begin{array}{c}
a(k)\\
B(k)
\end{array}\right)=0,
\end{equation}
where
\begin{equation}
M(k)=\left(\begin{array}{cc}
\frac{\di3}{\di2}J_c+J_c\cos k-J_c\Big(z(k)+\frac{\di1}{\di z(k)}\Big)\cos\frac{\di k}{\di2}-
J_{\bot}&iz(k)J_d\cos\frac{\di k}{\di2}\\
-\frac{\di iJ_d}{\di2}\cos\frac{\di
k}{\di2}&\Big(\frac{\di3}{\di4}J_c-\frac{\di
J_d}{\di2}\Big)z(k)-J_c\cos\frac{\di k}{\di2}
\end{array}\right).
\end{equation}

The Eq. (17) is solvable only for $det M(k)=0$, or,
\begin{eqnarray}
\Big[z^2(k)J_c\cos\frac{k}{2}+\Big(J_{\bot}-\frac{3}{2}J_c-J_c\cos{k}\Big)z(k)+
J_c\cos\frac{k}{2}\Big]&&\nonumber\\
\times\Big[\Big(\frac{3}{2}J_c-J_d\Big)z(k)-2J_c\cos\frac{k}{2}\Big]+z^2(k)J^2_d\cos^2\frac{k}{2}&=&0.
\end{eqnarray}

The Eq. (19) added by the conditions (14) and (16) completely
defines the coherent spectrum for $\hat H^{eff}$. The truncation
originates from a failure of any of the conditions (14) or (16).
For the first possibility the truncation wave vector $k_{trunc}$
coincides with the, critical wave vector $k_c$ defined as,
\begin{equation}
|z(k_c)|=1.
\end{equation}
For the second one it coincides with the branching wave vector
$k_b$ related to passing of solutions of the Eq. (19) into the
complex plane.

In order to clear the nature of the truncation point for an
arbitrary set of coupling parameters (however limited by (3) and
(5)) let us first examine the case when the condition (20) is
satisfied just at the branching point. In other words we are
interesting in $k_c=k_b=k_{bc}$ when the Eq. (19) has a
twice-degenerate solution $z(k_{bc})$ so that the same one has the
equation obtained from (19) by differentiating of its left side
with respect to $z(k)$. Using an auxiliary variable
$f=z(k_{bc})\cos\frac{k_{bc}}{2}$ and taking in mind that
according to (16) and (20) $z^2(k_{bc})=1$, we represent (at
$k=k_{bc}$) the Eq. (19) and its "derivative" equation as the
following system,
\begin{eqnarray}
4J_c^2f^3+(J_d^2+2J_dJ_c-7J_c^2)f^2+J_c(4J_c-2J_d-2J_{\bot})f+(\frac{3}{2}J_c-J_d)(J_{\bot}-\frac{J_c}{2})=0,&&\nonumber\\
2J_c^2f^3+(J_d^2+2J_dJ_c-5J_c^2)f^2+J_c(\frac{7}{2}J_c-2J_d-J_{\bot})f+(\frac{3}{2}J_c-J_d)(J_{\bot}-\frac{J_c}{2})=0,&&
\end{eqnarray}
which is solvable only for,
\begin{equation}
J_cJ_d(2J_{\bot}-J_c)(3J_c-2J_d)=0.
\end{equation}
(The left side of (22) was obtained from the resultant of the two
polynomials in the left sides of (21)). The solution $J_c=0$ of
the Eq. (22) is not interesting because in this case the Eq. (19)
is singular and solvable only for $k=\pi$. The solution
$J_c=2J_{\bot}$ is inconsistent with (5). The solution $3J_c=2J_d$
is artificial because in this case $\cos\frac{k}{2}$ factorizes
from the left side of (19), and therefore at $k=\pi$ the Eq. (19)
is identically satisfied for all $z(\pi)$. The solution $J_d=0$
relates to zero asymmetry when the corresponding truncation wave
vector $k_0$,
\begin{equation}
\cos\frac{k_0}{2}=\frac{1}{2}\Big(\sqrt{\frac{2J_{\bot}}{J_c}}-1\Big),
\end{equation}
may be easily obtained from the Eq. (19) which also gives,
\begin{equation}
z(k_0)=-1.
\end{equation}
The formula (23) has a clear physical interpretation. Really as it
follows from the results of the Refs. \cite{4} and \cite{5}
(related to {\it symmetric} spin ladders) at $k=k_0$ the "bare"
one-magnon branch with dispersion
$E^{magn}_{bare}(k)=J_{\bot}-\frac{3}{2}J_c+J_c\cos k$ intersects
the lower bound of the scattering two magnon continuum \cite{5},
\begin{equation}
E^{2magn,low}_{bare}(k)=2J_{\bot}-3J_c-2J_c\cos\frac{k}{2}.
\end{equation}
The above result confirm the general statement suggested in
\cite{6}-\cite{8} that even an extremely small asymmetry may
change drastically a magnon mode. As it follows from (23) at
$J_d\rightarrow0$ the truncation occurs only for $9J_c>2J_{\bot}$.
Since $J_{\bot}>0$ the parameter $J_c$ also must be positive.

In order to find a nature of the truncation at $J_d\neq0$ let us
study an evolution of $z(k_b)$ for small $J_d$. If the condition
(14) is satisfied for $k=k_b$ then the truncation originates from
branching and $k_{trunc}=k_b$. However in the opposite side for
$|z(k_b)|>1$ it will be $k_{trunc}=k_c$.

Taking for $J_d/J_c\ll1$ and $k\approx k_0$ the following
infinitesimal representation $z(k)=-1+\epsilon(k)$, using the
following notations $t(k)=\cos k/2$, $t_0=\cos k_0/2$ and the
formula,
\begin{equation}
\frac{J_{\bot}}{J_c}-\frac{3}{2}-\cos k=2(t_0^2+t_0-t^2(k)),
\end{equation}
which follows from (23) we obtain from (19) by omitting the term
$\epsilon^3(k)$ the following equation,
\begin{equation}
\alpha(k)\epsilon^2(k)+\beta(k)\epsilon(k)+\gamma(k)=0.
\end{equation}
Here
\begin{eqnarray}
\alpha(k)&=&1+\frac{t(k)}{\Delta_1}+2\frac{(t(k)-t_0)(t(k)+t_0+1)}{t(k)}-\frac{J_d^2t(k)}{2J_c^2\Delta_1},\nonumber\\
\beta(k)&=&2\frac{(t_0-t(k))(t(k)+t_0+1)}{t(k)}(2+\frac{t(k)}{\Delta_1})+\frac{J_d^2t(k)}{J_c^2\Delta_1},\nonumber\\
\gamma(k)&=&2\frac{(t(k)-t_0)(t(k)+t_0+1)}{t(k)}(1+\frac{t(k)}{\Delta_1})-\frac{J_d^2t(k)}{2J_c^2\Delta_1},
\end{eqnarray}
and $\Delta_1=3/4-J_d/(2J_c)$.

The branching wave vector is characterized by the following
condition,
\begin{equation}
D(k_b)=\beta^2(k_b)-4\alpha(k_b)\gamma(k_b)=0.
\end{equation}
After its linearization with respect to small parameters
$t(k)-t_0$ and $J^2_d/J^2_c$ this equation reduces at first to,
$\gamma(k_b)=0$ and then to,
\begin{equation}
\cos\frac{k_b}{2}\approx
t_0+\frac{J_d^2t_0^2}{4J_c^2(2t_0+1)(t_0+\Delta_1)}.
\end{equation}
According to (27) and (29),
$\epsilon(k_b)=-\beta(k_b)/(2\alpha(k_b))$, or using (30) and
(28),
\begin{equation}
\epsilon(k_b)\approx-\Big(\frac{J_dt_0}{2J_c(t_0+\Delta_1)}\Big)^2.
\end{equation}

Since $\epsilon(k_b)<0$ the condition (14) fails for $z(k_b)$.
Therefore for $J_d^2\ll J_c^2$ there must be,
\begin{equation}
k_{trunc}=k_c.
\end{equation}
Since for $J_d\neq0$ the wave vector $k_c$ evolve continuously
from $k_0$ the Eqs. (24), (20) and (16) give,
\begin{equation}
z(k_c)=-1.
\end{equation}
Despite the Eqs. (32) and (33) were proved for $J_d^2\ll J_c^2$
they are right for all $J_d$. Really if for some region of $J_d$
it will be $k_{trunc}=k_b$ then there must be a point where
$k_c=k_b$. But as it was shown above $k_0 $ is the only one point
of such type.

The Eqs. (15) and (33) give the following representation for the
magnon energy at the truncation point,
\begin{equation}
E_{trunc}=2J_{\bot}-3J_c-2J_c\cos\frac{k_c}{2}.
\end{equation}

The magnon branch approaches the bottom of the two-magnon
continuum tangentially,
\begin{equation}
\frac{\partial}{\partial
k}E^{2magn,low}_{bare}(k)\Big|_{k=k_c}=\frac{\partial}{\partial
k}E(k)\Big|_{k=k_c}=J_c\sin\frac{k_c}{2}.
\end{equation}
The Eq. (35) may be easily derived from (25) and (15) using an
auxiliary relation,
\begin{equation}
\frac{\partial}{\partial
k}\Big(z(k)+\frac{1}{z(k)}\Big)\Big|_{k=k_c}=0,
\end{equation}
which follows from (33). The same result was obtained in \cite{7}
by a different approach.

Let us notice that the singularity at $t_0+\Delta_1=0$ in the
formulas (30) and (31) originates from a resonance between the
one-magnon and bound two-magnon states. Really for $t_0=-\Delta_1$
the Eq. (19) has the thrice-degenerated solution related to both
these states. This special case is not considered in the present
paper.

\section{Magnon dispersion near the gap}
Let us turn to the opposite side of the spectrum related to
$k=\pi$. As it follows from (19) $z(k)$ is an odd function and
$z(\pi)=0$. Therefore for $k\approx\pi$ we may put,
\begin{equation}
z(k)\approx z_1(\pi-k)+z_3(\pi-k)^3.
\end{equation}
Then from (15) and (37) follows that for $z_1^3-z_1/12-z_3>0$ the
dispersion at $k\approx\pi$ takes the form,
\begin{equation}
E(k)\approx E_{gap}\Big(1+\frac{v_{spin}^2}{2E^2_{gap}}(\pi-k)^2\Big),
\end{equation}
where the gap energy $E_{gap}$ and the spin velocity $v_{spin}$ are given by
\begin{equation}
E_{gap}=2J_{\bot}-3J_c+\frac{J_c}{2z_1},\quad \frac{v_{spin}}{E_{gap}}=
\sqrt{\frac{J_c(z_1-\frac{\di1}{\di12z_1}-\frac{\di z_3}{\di z_1^2}\Big)}{2J_{\bot}-3J_c+\frac{\di J_c}{\di2z_1}}}.
\end{equation}
Since both $E_{gap}$ and $v_{spin}$ may be obtained by an
experiment \cite{6} we shall express them explicitly from the
coupling constants.

Substituting (37) into (19) we obtain the following system of
equations on the coefficients $z_1$ and $z_3$,
\begin{eqnarray}
\Big[\Big(J_{\bot}-\frac{J_c}{2}\Big)z_1+\frac{J_c}{2}\Big]\Big[\Big(\frac{3}{2}J_c-J_d\Big)z_1-J_c\Big]&=&0,\\
\Big[\Big(J_{\bot}-\frac{J_c}{2}\Big)z_1+\frac{J_c}{2}\Big]\Big[\Big(\frac{3}{2}J_c-J_d\Big)z_3+\frac{J_c}{12}\Big]
\nonumber\\
+\Big[\frac{J_c}{2}\Big(z_1^2-z_1-\frac{1}{12}\Big)+z_3\Big(J_{\bot}-\frac{J_c}{2}\Big)\Big]
\Big[\Big(\frac{3}{2}J_c-J_d\Big)z_1-J_c\Big]+\frac{J_d^2z_1^2}{4}&=&0.
\end{eqnarray}
The Eq. (40) has two solutions,
\begin{equation}
z_1^{magn}=-\frac{J_c}{2J_{\bot}-J_c},\quad
z_1^{bound}=\frac{2J_c}{3J_c-2J_d},
\end{equation}
related to magnon and bound two-magnon branches \cite{4},\cite{5}.
According to the first equation in (42)
$z_3^{magn}(J_{\bot}-J_c/2)=-J_cz_3^{magn}/(2z_1^{magn})$ so from
(41) follows,
\begin{equation}
z_1^{magn}-\frac{1}{12z_1^{magn}}-\frac{z_3^{magn}}{(z_1^{magn})^2}=1-\frac{J_d^2}{J_c(4J_{\bot}+J_c-2J_d)},
\end{equation}
and according to (39) one can obtain,
\begin{equation}
E_{gap}=J_{\bot}-\frac{5}{2}J_c,\quad
\frac{v_{spin}}{E_{gap}}=\sqrt{\frac{2J_c}{2J_{\bot}-5J_c}\Big(1-\frac{J_d^2}{J_c(4J_{\bot}+J_c-2J_d)}\Big)}.
\end{equation}
As it follows from (44) the point $k=\pi$ corresponds to an energy
minimum (the gap) only for $J_c(4J_{\bot}+J_c-2J_d)>J_d^2$.
(According to the comment after the Eq. (24) we suppose that
$J_c>0$.)

Using (33) and (34) we may represent the Eq. (19) in the point
$k=k_c$ as follows,
\begin{eqnarray}
&\Big(\frac{\di(E_{trunc}-2E_{gap})\cos^2\frac{\di k_c}{\di2}}{\di1-\cos\frac{\di k_c}{\di2}}+E_{gap}-E_{trunc}\Big)
&\nonumber\\
&\times\Big[\frac{\di(E_{trunc}-2E_{gap})\Big(3+4\cos\frac{\di
k_c}{\di2}\Big)}{\di4\Big(1-\cos\frac{\di k_c}{\di2}\Big)}
-J_d\Big]=J^2_d\cos^2\frac{\di k_c}{\di2}.&
\end{eqnarray}
where the parameters $J_{\bot}$ and $J_c$ are excluded by (34) and
(44). The Eq. (45) may be used for obtaining the parameter $J_d$
directly from an experimental data.

\section{One-magnon dynamical structure factor near the threshold}
We use the following representation for the dynamical structure factor (DSF),
\begin{equation}
S_{\alpha\beta}({\bf
q},\omega)=\lim_{N\rightarrow\infty}\frac{1}{N}\sum_{\mu}\langle0|{\bf
S}^{\alpha}({\bf q})|\mu\rangle\langle\mu|{\bf S}^{\beta}(-{\bf
q})|0\rangle\delta(\omega-E_{\mu}).
\end{equation}
Here ${\bf S}({\bf q})$ is the spin Fourier transformation
associated with the two dimensional vector ${\bf q}=(q,q_{rung})$
with leg and rung components. Since the latter has only two
possible values $0$ and $\pi$ we may study them separately,
\begin{equation}
{\bf S}(q,0)=\sum_n{\rm e}^{-iqn}({\bf S}_{1,n}+{\bf
S}_{2,n}),\quad {\bf S}(q,\pi)=\sum_n{\rm e}^{-iqn}({\bf
S}_{1,n}-{\bf S}_{2,n}).
\end{equation}

According to the following pair of relations, ${\bf
S}(q,0)|0\rangle=0$, ${\bf S}(-q,\pi)|0\rangle=\sum_n{\rm
e}^{iqn}...|1\rangle_n...$, we may reduce the matrix elements in
(46)
\begin{equation}
\langle\mu|{\bf S}(q,0)|0\rangle=0,\quad
^{\alpha}\langle k|{\bf S}^{\beta}(-q,\pi)|0\rangle=\delta_{\alpha\beta}\delta_{kq}
\frac{\sqrt{N}a(q)}{Z(q)},
\end{equation}
therefore, the DSF has purely diagonal form,
$S_{\alpha\beta}(q,\pi,\omega)=\delta_{\alpha\beta}S(q,\pi,\omega)$,
while the one-magnon contribution is purely coherent,
\begin{equation}
S_{magn}(q,\pi,\omega)=A_{magn}(q)\delta(\omega-E^{magn}(q)),
\end{equation}
where
\begin{equation}
A_{magn}(q)=\left|\frac{a_{magn}(q)}{Z_{magn}(q)}\right|^2.
\end{equation}

According to (8) and (13),
\begin{equation}
Z_{magn}(k)=\sqrt{|a_{magn}(k)|^2+\frac{2|B_{magn}(k)|^2z^2(k)}{1-z^2(k)}}.
\end{equation}
For $q\rightarrow k_c$, it will be $A_{magn}(q)\propto1-z^2(q)$,
so as it follows from (33) $A_{magn}(k_c)=0$. The same result was
obtained in \cite{7} by different approach.

Finitely let us notice that a rather similar effect of
hybridization between magnon and phonon modes was studied in
\cite{10}. However in the latter case a magnon mode does not
truncate (because there is no decay) and therefore the
corresponding structure factor does not turn to zero.

\section{Summary and discussion}
In this paper for a rung-dimerized asymmetric spin ladder we
suggested an effective model which neglects all states with $n>2$
bare magnons. Using Bethe Ansatze we studied the effect of magnon
mode truncation resulting from magnon decay and clarified its
mathematical nature (see the Eq. (32)). We obtained the four
equations (see (34), (44) and (45)) coupling the interactions
constants of our model (namely $J_{\bot}$, $J_c$ and $J_d$) with
the truncation wave vector, gap and truncation energies and spin
velocity.

Of course the neglect of the states with $n>2$ bare magnons is a
rather rough approximation. Really an intersection between the
one- and two-magnon scattering modes is possible only for a wide
band system. In this case the bare $n>2$ zones also lie not so far
from the magnon mode and therefore give a rather essential
contribution to it. However if we concern only on the gap and
truncation points then our model produces a good approximation.
Really near the gap the magnon energy is minimal and lies far
below the bare $n>2$ magnon modes. For example as it follows from
(44) for $J_d\ll J_{\bot},J_c$ even the $n=2$ correction is small.
From the other side since the $1\rightarrow2$ decay threshold lies
on a {\it finite} distance below the $1\rightarrow3$ one the
latter is not sufficient at the vicinity of the truncation point
where the parameters $(E(k_c)-E(k))/E_{gap}$ and
$(E(k_c)-E(k))/(E_{trunc}-E_{gap})$ are small. Therefore the
infinitesimal analysis of the Sect. 3 (Eqs. (27)-(32)) gives the
right picture of the truncation (the Eq. (33)).

There is only one known asymmetric rung-dimerized spin ladder
compound namely the CuHpCl (see \cite{11} and references therein).
However the effect of truncation was not observed in this
material. This fact is clear because the gap energy in CuHpCl (0.9
mev) is bigger than the magnon bandwidth (0.5 mev) so the magnon
mode does not intersect with the scattering two magnon continuum.

Despite none wide-band asymmetric rung-dimerized spin ladder
compound was found up to now we suppose that this may likely
happen in not so remote future. Then the results of our paper
probably will be useful for a theoretical study of such compound.

The author is grateful to S.~L.~Ginzburg, S.~V.~Maleyev and
A.~V.~Syromyatnikov for the interest and helpful discussion.

\end{document}